\documentclass[11pt]{article}

\usepackage[preprint]{latex/acl}

\usepackage{times}
\usepackage{latexsym}

\usepackage{hyperref}
\usepackage{url}
\usepackage{pifont}
\usepackage[utf8]{inputenc}
\usepackage[T1]{fontenc}
\usepackage{booktabs}
\usepackage{amsfonts}
\usepackage{nicefrac}
\usepackage{microtype}
\usepackage{lipsum}
\usepackage{graphicx}
\usepackage[table]{xcolor}
\usepackage{amsmath}
\usepackage{amssymb}
\usepackage{mathtools}
\usepackage{amsthm}
\usepackage{algorithm}
\usepackage{algorithmic}
\usepackage{natbib}
\usepackage{xcolor}
\usepackage{subcaption}
\usepackage{wrapfig}
\usepackage{multirow}
\usepackage{tablefootnote}
\usepackage{array}

\usepackage[T1]{fontenc}

\usepackage[utf8]{inputenc}

\usepackage{microtype}

\usepackage{inconsolata}

\usepackage{graphicx}

\title{MuseAgent-1: Interactive Grounded Multimodal Understanding of Music Scores and Performance Audio}

\author{
Qihao Zhao\textsuperscript{1}\thanks{Contributed equally.},
Yunqi Cao\textsuperscript{2}\footnotemark[1],
Yangyu Huang\textsuperscript{3},
Hui Yi Leong\textsuperscript{4},
Fan Zhang\textsuperscript{2},
Kim-Hui Yap\textsuperscript{1},
Wei Hu\textsuperscript{2}\thanks{E-mail: \href{mailto:huwei@mail.buct.edu.cn}{huwei@mail.buct.edu.cn}}
\\[0.5em]
\textsuperscript{1}Nanyang Technological University\\
\textsuperscript{2}Beijing University of Chemical Technology\\
\textsuperscript{3}Microsoft\\
\textsuperscript{4}University of Chicago
}

\begin{document}
\maketitle

\begin{abstract}
Despite recent advances in multimodal large language models (MLLMs), their ability to understand and interact with music remains limited. Music understanding requires grounded reasoning over symbolic scores and expressive performance audio, which general-purpose MLLMs often fail to handle due to insufficient perceptual grounding.
We introduce MuseAgent, a music-centric multimodal agent that augments language models with structured symbolic representations derived from sheet music images and performance audio. By integrating optical music recognition and automatic music transcription modules, MuseAgent enables multi-step reasoning and interaction over fine-grained musical content.
To systematically evaluate music understanding capabilities, we further propose MuseBench, a benchmark covering music theory reasoning, score interpretation, and performance-level analysis across text, image, and audio modalities. Experiments show that existing MLLMs perform poorly on these tasks, while MuseAgent achieves substantial improvements, highlighting the importance of structured multimodal grounding for interactive music understanding.

\end{abstract}

\section{Introduction}

\begin{figure*}[ht]
\small
\centering
\includegraphics[width=2\columnwidth]{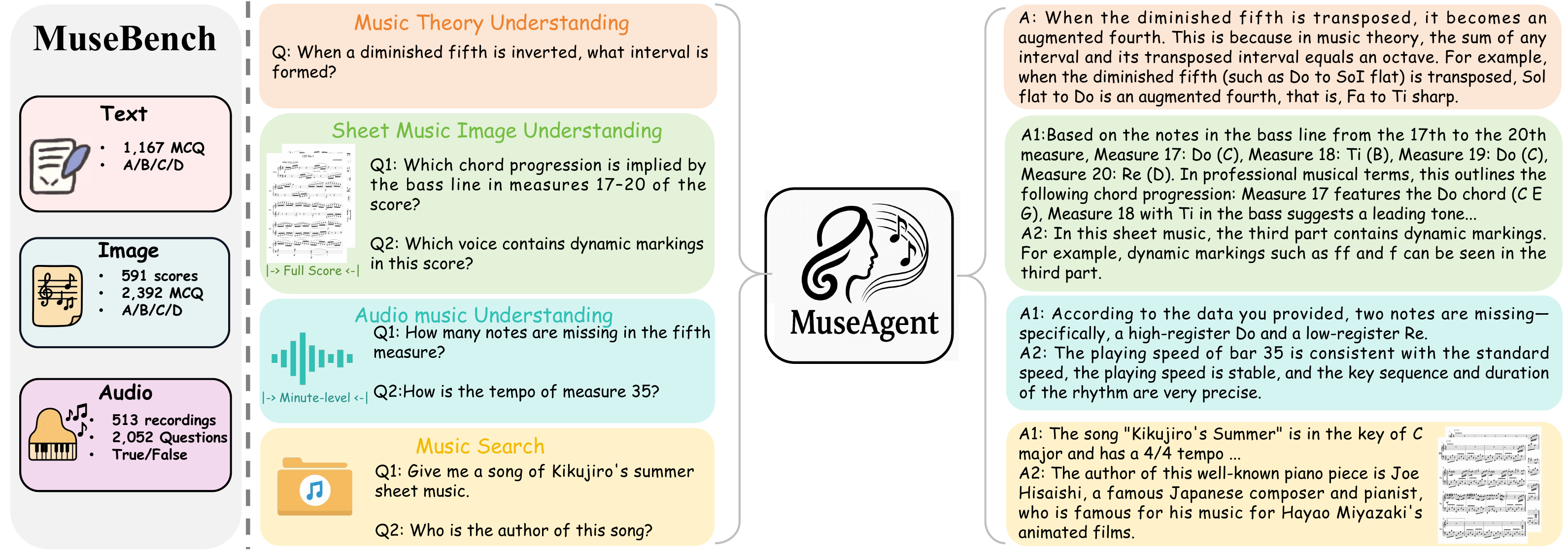}
\caption{Overview of \textbf{MuseBench} and \textbf{MuseAgent}. MuseBench consists of multimodal music understanding tasks across text, image, and audio modalities, covering music theory, sheet score analysis, performance interpretation. MuseAgent integrates these modalities via sheet symbolic recognition, audio alignment, and retrieval modules, enabling large language models to answer complex music questions.}
\label{Fig1}

\end{figure*}

\begin{flushright}
\small\itshape
“O Muses, O high genius, now help me!”\\
\upshape--- Dante, \textit{Inferno}, Canto II, line 7
\end{flushright}
\vspace{-5px}
Music is a structured yet expressive domain, making it a compelling testbed for artificial intelligence. It spans symbolic representations such as notated scores and expressive acoustic realizations such as performances, each requiring distinct perceptual and reasoning capabilities. As both a formal system and an emotional medium, music challenges AI models to reason across modalities with high precision and contextual nuance \cite{essid2012music, correard2021crossmodal}. These requirements expose fundamental limitations of current multimodal systems, particularly large language models that lack structured grounding in symbolic and temporal representations.

Among musical forms, piano music occupies a uniquely central role in the Western repertoire, supported by centuries of standardized notation and extensive performance archives. Its wide pitch range, dense polyphony, and two-hand coordination make piano music a particularly demanding stress test for multimodal reasoning systems \cite{hawthorne2018enabling}. Understanding piano music requires jointly modeling complex symbolic structures—such as pitch, rhythm, and dynamics—and expressive performance attributes including timing, articulation, and rubato. This integration is essential for interactive applications such as transcription, accompaniment, digital archiving, and music education \cite{benetos2019automatic}.

Despite significant progress in isolated tasks, existing approaches remain insufficient for holistic and interactive music understanding. Prior work has advanced single-modality problems such as Optical Music Recognition (OMR) for scores and Automatic Music Transcription (AMT) for audio. Large-scale datasets like MAESTRO \cite{hawthorne2018enabling} and generative models such as MusicLM \cite{agostinelli2023musiclm} demonstrate the feasibility of modeling symbolic and acoustic music representations, yet they fall short of enabling fine-grained cross-modal reasoning. Recent Multimodal Large Language Models (MLLMs), including GPT-4o \cite{openai2023gpt4} and Gemini \cite{google2023gemini}, promise general cross-modal capabilities but perform poorly on music understanding tasks that require precise symbolic parsing or long-form performance analysis. This failure stems largely from the lack of domain-specific perceptual grounding and the inability to align symbolic and acoustic modalities at high temporal resolution.

Several recent systems have attempted to bridge this gap by augmenting LLMs with external tools. AudioGPT \cite{huang2024audiogpt} and MusicAgent \cite{yu2023musicagent} represent early efforts toward music-aware agents by coupling language models with domain-specific components. However, these systems often struggle with complex notation, long-duration performance recordings, and structured interaction, and they lack systematic evaluation protocols tailored to music understanding. Meanwhile, advances in Retrieval-Augmented Generation (RAG) \cite{lewis2020rag, guu2020realm, borgeaud2022retro} demonstrate that grounding language models in structured external representations can substantially improve domain-specific reasoning and reduce hallucination. Existing multimodal RAG frameworks \cite{shuster2022multimodal, luo2023mmrag, liu2023visualrag}, however, primarily focus on text–vision–speech settings and rarely address music, where symbolic scores and expressive audio must be tightly aligned.

To address these challenges, we propose MuseAgent, a multimodal retrieval-augmented agent designed specifically for music understanding and interaction. MuseAgent integrates a large language model with specialized perceptual front-ends: (i) a measure-wise OMR module that converts sheet music images into symbolic representations such as ABC notation \cite{yuan-etal-2024-chatmusician}, (ii) an AMT-based performance analysis module that aligns audio recordings with MusicXML scores and extracts expressive features in structured JSON formats, and (iii) a retrieval module that enables both explicit and implicit access to symbolic and audio libraries. These components ground the language model in structured multimodal representations, while a memory bank supports long-context, multi-turn reasoning over musical content.

Evaluating such music understanding agents poses a non-trivial challenge. Existing datasets and benchmarks focus primarily on transcription or generation, and do not assess an agent’s ability to reason interactively across symbolic scores and performance audio. To enable systematic evaluation, we introduce MuseBench, the first benchmark designed to assess multimodal music understanding agents. Centered on piano repertoire, MuseBench includes tasks such as score–audio alignment, performance error detection, and expressive deviation analysis, repurposing resources like MAESTRO into high-level reasoning tasks suitable for evaluating agentic music understanding.

Our experiments on MuseBench reveal that general-purpose MLLMs exhibit limited capabilities in fine-grained symbolic and performance-level music reasoning. In contrast, MuseAgent demonstrates substantial improvements on challenging image- and audio-based tasks, validating the effectiveness of structured perceptual grounding and agent-based multimodal reasoning for interactive music understanding.

\section{Related Work}

\paragraph{Music Understanding Systems and Agents.}
Recent advances in general-purpose agents, such as ReAct, Auto-GPT, and Gorilla, have demonstrated the effectiveness of tool-augmented reasoning for complex tasks. In the music domain, early systems such as MuseNet~\cite{payne2019musenet} and MusicLM~\cite{agostinelli2023musiclm} primarily focus on symbolic or audio music generation, rather than music understanding or interactive reasoning. More recent music-oriented multimodal language models adopt domain-specific training strategies. For example, MusiLingo~\cite{deng2024musilingo} targets music captioning and question answering via instruction-tuned audio--language modeling, MuMu-LLaMA~\cite{liu2024mumullama} unifies music audio, images, and language within a single multimodal framework, and NotaGPT~\cite{tang2025notamultimodalmusicnotation} focuses on music notation understanding through a vision--language model. SymphonyNet~\cite{liu2022symphonynet} further demonstrates symbolic music generation at orchestral scale.

While these domain-specific models highlight the value of multimodal alignment and specialized training, they are typically designed as monolithic architectures and do not explicitly support structured interaction, tool invocation, or multi-step reasoning. In parallel, agent-based systems such as MusicAgent~\cite{yu2023musicagent} and AudioGPT~\cite{huang2024audiogpt} begin to couple large language models with domain-specific tools for processing music inputs. Although promising, these systems are not designed for fine-grained music understanding that requires jointly reasoning over symbolic scores and expressive performance audio, nor do they target complex piano repertoire. In contrast, our proposed \textit{MuseAgent} advances this line of work by introducing a music-centric agent architecture that integrates structured perceptual grounding, retrieval-based reasoning, and agentic orchestration across symbolic and acoustic modalities.

\paragraph{Music Understanding Benchmarks.}
A critical challenge in developing music understanding agents lies in evaluation. Existing music datasets, such as MAESTRO~\cite{hawthorne2019maestro} and URMP~\cite{li2018urmp}, provide aligned score--audio pairs and have been widely used for transcription, alignment, and generation tasks. However, these resources lack task-oriented evaluation protocols that assess high-level reasoning or interactive understanding. Recent multimodal question-answering datasets, including MUSIC-AVQA~\cite{li2022musicavqa} and MuChoMusic~\cite{weck2024muchomusic}, extend music benchmarks toward Q\&A settings, but do not emphasize detailed symbolic score interpretation or nuanced performance-level reasoning. In contrast, our proposed \textit{MuseBench} is designed explicitly to evaluate music understanding agents, assessing their ability to reason jointly over text, score images, and performance audio across a unified set of piano-centric tasks.

\paragraph{General Multimodal Language Models.}
General-purpose multimodal language models, such as GPT-4o~\cite{openai2023gpt4}, Gemini~\cite{anil2023gemini}, Qwen~\cite{qwen2.5}, and LLaVA~\cite{liu2023llava}, achieve strong performance on text--vision benchmarks. However, recent studies~\cite{weck2024muchomusic} demonstrate that these models struggle with music understanding tasks, particularly when confronted with structured musical notation or expressive performance audio. In the absence of explicit symbolic grounding, such models often rely on linguistic priors or hallucinate musical content. These limitations motivate the need for music-aware agents that integrate domain-specific perception and structured reasoning, as exemplified by our MuseAgent.

\paragraph{Distinctiveness of Our Work.}
In summary, prior research on music understanding has largely focused on isolated modalities, monolithic domain models, or generation-oriented objectives, while existing benchmarks lack the ability to evaluate interactive, agent-based reasoning over music. Our work uniquely combines a music-centric multimodal agent with an agent-oriented benchmark. MuseAgent enables structured reasoning over both symbolic scores and expressive performance audio, while MuseBench provides a unified evaluation framework tailored to assessing such capabilities. Together, they establish a new foundation for studying fine-grained, interactive multimodal music understanding.

\section{MuseAgent}
\label{PianoMuse}
\begin{figure*}[ht]
\small
\centering
\includegraphics[width=2\columnwidth]{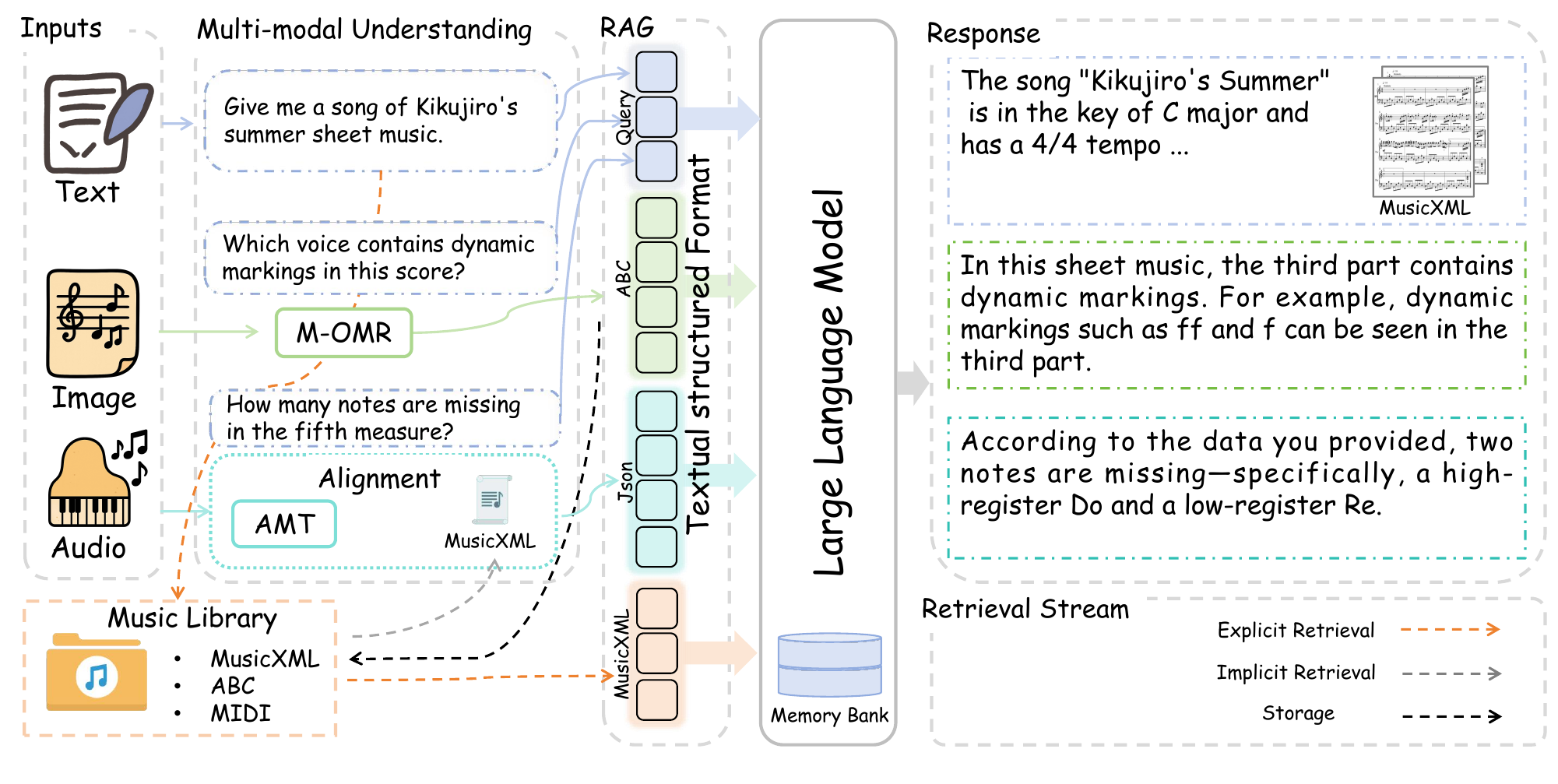}
\caption{The MuseAgent framework integrates M-OMR, AMT, and music retrieval (explicit/implicit) into a unified large language model (LLM)-based system. Each perceptual module converts raw multimodal inputs into structured symbolic representations (e.g., ABC, MusicXML, JSON), which are incorporated into a retrieval-augmented generation (RAG) pipeline. The LLM acts as an agentic controller that dynamically orchestrates module usage depending on user intent, while a memory bank supports multi-turn dialogue and retrieval of prior outputs. }
\label{Fig2}
\vspace{-5px}
\end{figure*}

We propose \textbf{MuseAgent}, a multimodal retrieval-augmented agent designed for structured and interactive music understanding. Unlike fixed perception–reasoning pipelines, MuseAgent adopts an \emph{agentic orchestration loop} in which a large language model dynamically coordinates domain-specific perceptual modules, retrieval operations, and symbolic reasoning based on user intent. This design enables MuseAgent to perform multi-step reasoning over symbolic scores and expressive performance audio, addressing the challenges posed by agent-oriented music understanding tasks such as those defined in MuseBench.

As illustrated in Figure~\ref{Fig2}, MuseAgent integrates three core components: (i) perceptual grounding modules that convert raw multimodal inputs into structured symbolic representations, (ii) a retrieval-augmented generation (RAG) mechanism that grounds reasoning in external music knowledge, and (iii) a memory bank that supports long-context, multi-turn interaction.

\subsection{Agentic Workflow}

At the core of MuseAgent is an LLM-based controller that interprets user queries and dynamically determines which perceptual or retrieval modules to invoke. For example, queries concerning harmonic structure or notation trigger the measure-wise OMR module, while questions about timing, tempo stability, or performance accuracy invoke AMT-based alignment. Retrieval may be initiated explicitly by user requests or implicitly by the agent when additional symbolic context is required. This agentic workflow enables flexible, intent-driven reasoning across modalities rather than static, pre-defined processing pipelines.

\subsection{Measure-wise Optical Music Recognition}
\begin{figure}[ht]
\small
\centering
\includegraphics[width=0.99\columnwidth]{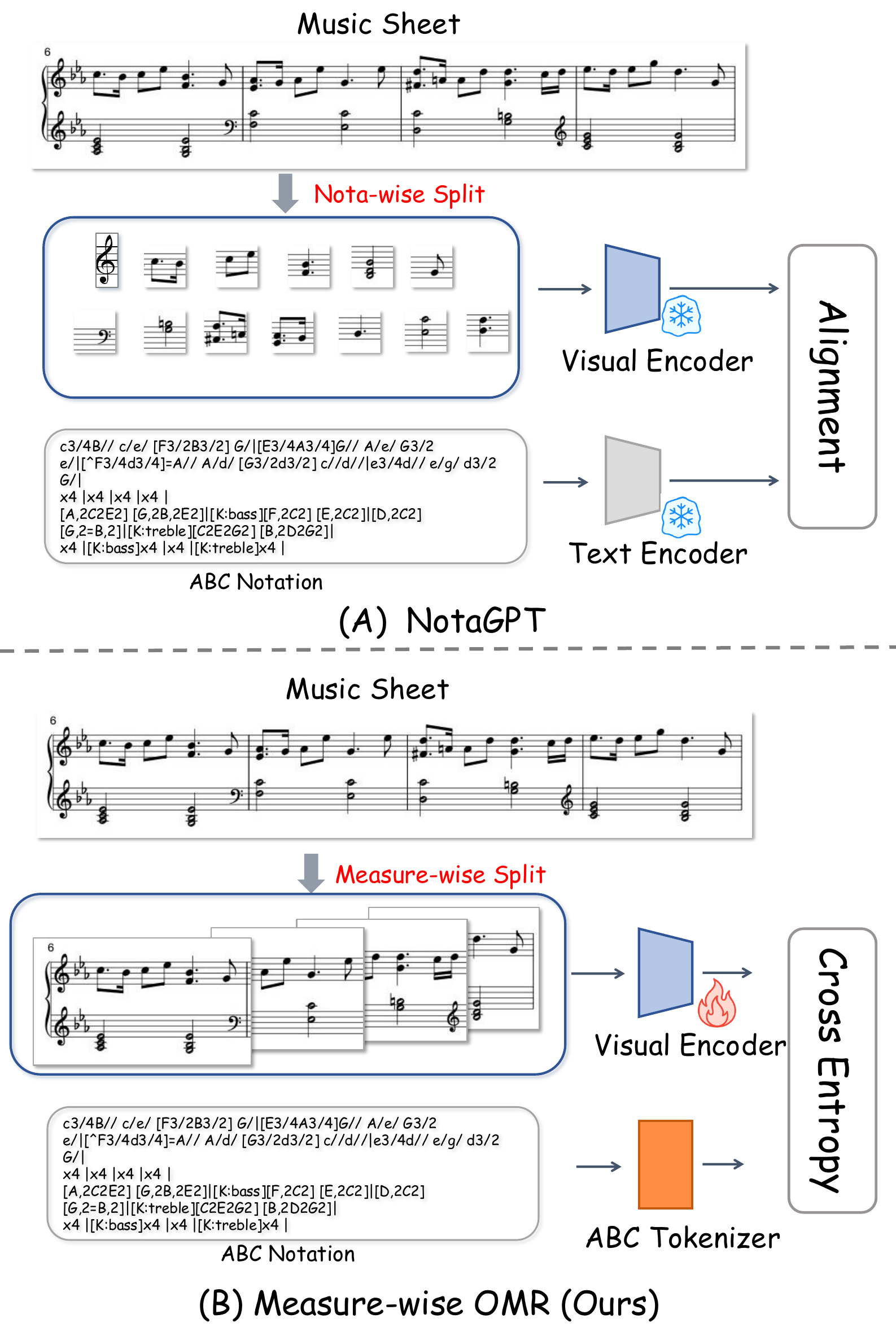}
\caption{Comparison between (A) \textit{NotaGPT}, which performs note-level segmentation with frozen visual and text encoders, and (B) our proposed \textit{Measure-wise OMR} approach. The flame symbol denotes trainable modules, while the snowflake symbol indicates frozen components.}
\label{Fig3}
\vspace{-10px}
\end{figure}
A key challenge for multimodal large language models (MLLMs) in music understanding is the modality gap between high-resolution score images and the symbolic reasoning required for musical analysis. Unlike natural images, music scores are densely structured and domain-specific, encoding hierarchical elements such as pitch, rhythm, and dynamics that general vision-language models struggle to interpret directly.

To address the challenge of structured music score recognition, we propose a Measure-wise Optical Music Recognition (M-OMR) module based on a ``divide-and-combine" strategy as a perceptual grounding module. The input sheet image is first divided into individual measures using visual layout cues such as staff lines and barlines. Each measure is treated as an independent visual unit, encoded into symbolic representation through localized recognition. These measure-level outputs are then combined to reconstruct the complete musical piece. The final result is expressed in ABC notation, a compact and structured symbolic format well-suited for downstream language model processing.

Unlike prior OMR approach, NotaGPT~\cite{tang2025notamultimodalmusicnotation}, which split score images at the note level and rely on frozen vision and text encoders, we adopt a measure-wise segmentation strategy. The different between them as shown in Figure~\ref{Fig3}. By treating each measure as a semantic unit, our model preserves musical structure and reduces noise from overly granular splitting. We train a ResNet-based~\cite{he2016deep} visual encoder, jointly trained with an LSTM~\cite{yu2019review} over the measure sequence to capture intra-score dependencies. Furthermore, we introduce a custom-designed ABC tokenizer tailored for ABC-notation representation. This tokenizer captures over hundreds of ABC-notation variants of music-specific constructs (e.g., key, meter, chords), producing more compact and structurally meaningful token sequences compared to general-purpose text encoders. More details are shown in Appendix~\ref{appendix:omr-examples}.

\subsection{AMT and Alignment}

To understand expressive performance audio, MuseAgent incorporates an Automatic Music Transcription (AMT) module and an audio-to-score alignment component as the audio perceptual grounding module. The AMT module transcribes raw audio into a symbolic representation (e.g., MusicXML) by extracting time-frequency features via the Constant-Q Transform \cite{schorkhuber2010constant}, and applying neural transcription models.

The resulting symbolic sequence is temporally aligned to a reference score using a hierarchical Hidden Markov Model (H-HMM) \cite{nakamura2015real}, which is robust to expressive timing variations, ornaments, and structural deviations such as repeats or skips. The alignment process produces structured outputs in JSON format, capturing onset timings, note correspondences, and expressive parameters.

These alignment outputs are then fused with user prompts and MusicXML files retrieved implicitly from the music library, forming the input to a retrieval-augmented generation (RAG) module. The RAG component composes these multimodal elements into an enriched prompt, enabling the language model to reason over both symbolic and auditory performance data. Implementation details, including model architecture and training configurations, are provided in Appendix~\ref{appendix:amt}.

\subsection{Music Retrieval Module}

The MuseAgent supports both explicit and implicit retrieval from a large-scale symbolic music library in formats such as ABC, MusicXML, and MIDI.

For \textbf{explicit retrieval}, users can issue direct natural language queries (e.g., “Give me a song of \emph{Kikujiro's Summer}”) to fetch matching scores. For \textbf{implicit retrieval}, the system performs internal searches conditioned on audio, and sheet context, selecting relevant symbolic files (e.g., auio-paired MusicXML) to be integrated into the the RAG pipeline.

Unlike traditional information retrieval methods, retrieval here is embedded into the agent loop: the LLM may explicitly respond to user queries or implicitly call the retrieval API to ground its reasoning. This design realizes agentic RAG for multimodal music.

\subsection{Music Theory Understanding and Dialogue Context}

In addition to the aforementioned capabilities, MuseAgent harnesses the intrinsic musical knowledge embedded in large language models to answer music-theoretical questions (e.g., “What interval results from inverting a diminished fifth?” or “Which mode begins on E in the C major scale?”). The effectiveness of this ability may vary across used LLMs. For evaluations of music theory understanding in different LLMs, please refer to Sec.~\ref{Results}.

The MuseAgent also supports memory capabilities for multi-turn conversations, for which we maintain a lightweight \textbf{memory bank} that stores intermediate module outputs, retrieved files, and previous model responses. The memory bank not only supports multi-turn reasoning, but also enables retrieval of prior structured outputs.

\section{MuseBench}

To enable systematic evaluation of \emph{music understanding agents}, we introduce \textbf{MuseBench}, a multimodal benchmark designed to assess structured and interactive reasoning over music. MuseBench targets agents that integrate perception, retrieval, and language-based reasoning, and evaluates their ability to jointly reason over \textbf{text}, \textbf{sheet music images}, and \textbf{performance audio}. Rather than measuring isolated recognition or generation skills, MuseBench focuses on high-level music understanding tasks that require symbolic grounding, cross-modal alignment, and multi-step reasoning—capabilities central to agent-based frameworks such as MuseAgent.

\subsection{Data Sources}

To support agent-oriented evaluation, MuseBench is constructed from high-quality symbolic, visual, and acoustic music data. The benchmark covers a wide range of styles, eras, and difficulty levels, including Baroque, Classical, Romantic, and contemporary piano repertoire. An overview of the dataset is shown in Figure~\ref{Fig1}, with further source details provided in Appendix~\ref{app:data_details}.

\subsection{Dataset Construction}

\subsubsection{Preprocessing}

We initially collected $\sim$3,000 candidate scores from multiple open repositories (see Appendix~\ref{app:data_details}). 
Scores were filtered for completeness, readability, and resolution quality. 
After normalization (resolution adjustment, background noise removal, and staff-line correction), $\sim$600 sheet images were retained. 
For audio, we collected 513 high-quality performance recordings. 
Each audio file was standardized to a uniform sampling rate and post-processed (denoising, normalization) to ensure clarity. 
To avoid copyright infringement, only recordings distributed under public licenses or explicitly provided by musicians with written consent were included.

\subsubsection{Annotation}

Each sheet music image was paired with an ABC-format symbolic file containing metadata such as title, composer, key, time signature, note durations, and rhythm. 
Expert musicians further annotated technical difficulty and performance-related elements. 
Each piece was aligned with professional piano audio recordings and converted into MusicXML with bar-level score–audio alignment, forming a standardized metadata pool for subsequent task construction. 
All annotations were performed by trained musicians with at least five years of formal music education. 
To ensure reliability, multiple annotators cross-validated the labels, achieving a Cohen’s $\kappa$ of 0.87.

\subsubsection{Definition}
\begin{figure}[ht]
\small
\centering
\includegraphics[width=0.99\columnwidth]{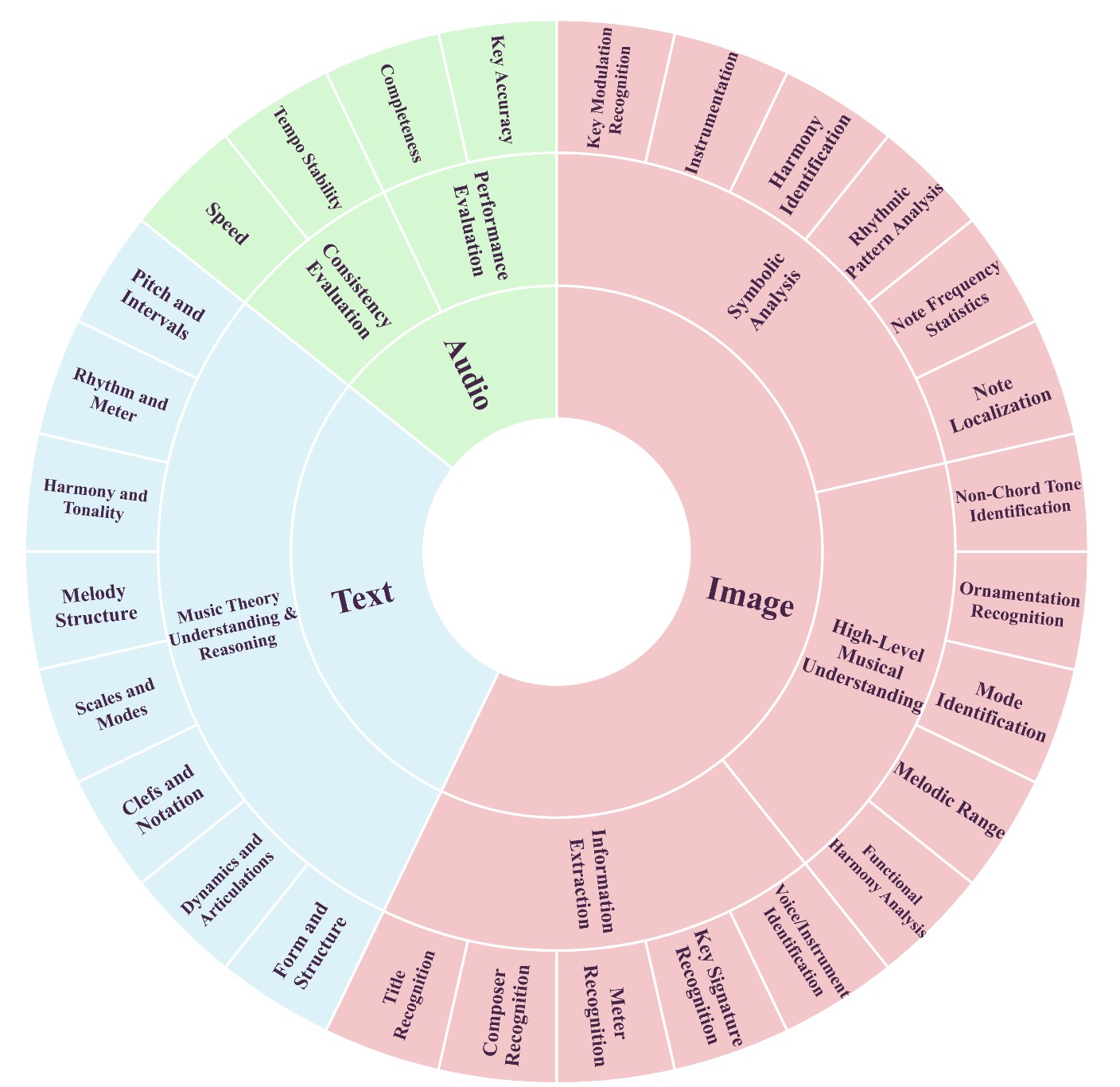}
\caption{\small Distribution of questions in MuseBench. It consists of 28 task types across three modalities. Tasks are relatively evenly distributed to ensure balanced evaluation.}
\label{Fig4}

\end{figure}

To evaluate the capabilities of MLLMs in music understanding and reasoning, we construct \textbf{MuseBench}, a benchmark collaboratively designed with expert musicians. It is structured around three core dimensions: (1) \textbf{music theory understanding}, (2) \textbf{sheet music understanding}, and (3) \textbf{performance audio analysis}. Task definitions and evaluation criteria were established under expert consensus to ensure relevance and rigor.

A detailed description of each sub-task and its design rationale is provided in Appendix~\ref{appendix:task-details}. 
In total, as shown in Figure 4, we define \textbf{28 specific tasks} across the three modalities and six sub-dimensions, ensuring a balanced distribution of question–answer pairs and task formats. Our design draws inspiration from evaluation frameworks such as MMMU~\cite{yue2024mmmu} and OmniBench~\cite{li2024omnibench}, while explicitly accounting for task difficulty and modality diversity to enable robust benchmarking.

\section{Experiment}

\subsection{Comparison on MuseBench}
\subsubsection{Baseline}
We evaluate MuseAgent on MuseBench against 17 representative multimodal language models to examine the effectiveness of agent-based grounding for music understanding. The selected baselines span a wide range of model paradigms, including: 
(i) \textbf{general-purpose language models} such as GPT-4.1, GLM-4, and Phi-4; 
(ii) \textbf{omni-modal models} including GPT-4o, Gemini 2.5-Pro, and Qwen2.5-Omni; 
(iii) \textbf{vision–language models} such as LLaVA, VisualGLM, and Qwen2.5-VL (7B/32B/72B); 
(iv) \textbf{audio-capable models} including Qwen2-audio and MuMu-LLaMA; and 
(v) \textbf{music-specialized models} such as NotaGPT for music notation understanding.

While these models differ in modality coverage and domain specialization, they are primarily designed as monolithic models without explicit agentic orchestration, structured perceptual grounding, or multi-step retrieval-based reasoning. This diverse set of baselines allows us to systematically assess the advantages of MuseAgent’s agent-based design over non-agent and weakly grounded multimodal systems on fine-grained music understanding tasks.

\subsubsection{Results Analysis}
\label{Results}
\begin{table}[t]
\centering
\scriptsize
\caption{MuseBench performance comparison. MuseAgent significantly improves image and audio understanding over omni-modal and music-specialized baselines.}
\label{tab:main-results}
    \begin{tabular}{l|l|c}
    \toprule
    Modality & Model & Accuracy (\%) \\
    \midrule
    \multirow{10}{*}{Text} 
        & GPT-4.1~\cite{openai2025gpt41} & \textbf{86.7} \\
        & GPT-4o~\cite{openai2023gpt4} &85.5 \\
        & Gemini2.5-Pro~\cite{comanici2025gemini} & 83.9\\
        & GPT-4.1-mini~\cite{openai2025gpt41} & 80.5 \\
        & GPT-4.1-nano~\cite{openai2025gpt41} & 73.3 \\
        & GLM-4-PLUS~\cite{glm2024chatglm} & 78.2 \\
        & GLM-4-FlashX~\cite{glm2024chatglm} & 60.3 \\
        & Qwen2.5-72B~\cite{qwen2.5} & 81.4 \\
        & Qwen2.5-32B~\cite{qwen2.5} & 79.8 \\
        & Qwen2.5-7B~\cite{qwen2.5} & 71.3 \\
        & Qwen2.5-Omni-7B~\cite{xu2025qwen2omni} & 63.3 \\
        & Phi-4-14B~\cite{abdin2024phi4technicalreport} & 67.2 \\
        & Random & 25.0 \\
    \midrule
    \multirow{8}{*}{Audio} 
        & MuseAgent (w/ GPT-4.1) & \textbf{79.1} \\
        & MuseAgent (w/ GPT-4o-mini) & 78.9 \\
        & MuseAgent (w/ GLM-4-FlashX) & 77.2 \\
        & MuseAgent (w/ GPT-4.1-Nano) & 63.9 \\
        & GPT-4o~\cite{openai2024gpt4omini} & 55.9 \\
        & Gemini2.5-Pro~\cite{comanici2025gemini} & 53.1 \\
        & MuMu-LLaMA~\cite{liu2024mumullama} & 51.7 \\
        & Qwen2-audio~\cite{qwen2.5} & 51.4 \\
        & Qwen2.5-Omni-7B~\cite{xu2025qwen2omni} & 50.6 \\
        & Random & 50.0 \\
    \midrule
    \multirow{8}{*}{Image}
        & MuseAgent (w/ GPT-4.1) & \textbf{74.1} \\
        & MuseAgent (w/ GLM-4-FlashX) & 72.7 \\
        & NotaGPT-7B~\cite{tang2025notamultimodalmusicnotation} & 68.1\\
        & GPT-4.1~\cite{openai2025gpt41} & 66.1 \\
        & GPT-4o~\cite{openai2024gpt4omini}  & 64.2\\
        & GPT-4.1-mini~\cite{openai2025gpt41} & 54.8 \\
        & Gemini2.5-Pro~\cite{comanici2025gemini} & 62.1 \\
        & Qwen2.5-VL-72B~\cite{qwen2.5} & 58.9\\
        & Qwen2.5-VL-32B~\cite{qwen2.5} & 55.7 \\
        & Qwen2.5-Omni-7B~\cite{xu2025qwen2omni}	& 44.6 \\
        & LLaVA-v1.5-13B~\cite{liu2023llava} & 38.9 \\
        & GLM-4V-9B~\cite{glm2024chatglm} & 37.1 \\
        & Random & 25.0 \\
    \bottomrule
    \end{tabular}
\end{table}
\textbf{Text Modality.} 
On purely textual tasks, general-purpose language models perform strongly. GPT-4.1 achieves the highest accuracy (86.7\%), followed closely by GPT-4o (85.5\%) and Gemini 2.5-Pro (83.9\%). Scaling model size yields moderate gains, as seen in Qwen2.5-72B (81.4\%) and Qwen2.5-32B (79.8\%), indicating that music theory questions largely align with the native capabilities of large language models.

\textbf{Audio Modality.} 
In contrast, general-purpose omni-modal models struggle with performance-level audio reasoning. GPT-4o (55.9\%) and Gemini 2.5-Pro (53.1\%) perform only marginally above random, while Qwen2.5-Omni (50.6\%) and the audio-specialized Qwen2-Audio (51.4\%) show similarly limited performance. These results suggest that raw audio inputs alone are insufficient for fine-grained music understanding. By explicitly incorporating AMT and score–audio alignment, MuseAgent equipped with GPT-4.1 achieves 79.1\% accuracy, demonstrating that structured perceptual grounding is essential for reasoning about expressive performance.

\textbf{Image Modality.} 
Vision–language models exhibit substantial difficulty in interpreting symbolic music notation. Models such as LLaVA (38.9\%) and GLM-4V (37.1\%) perform poorly, reflecting the challenges posed by dense and domain-specific score representations. Larger omni-modal models, including GPT-4o (64.2\%) and Gemini 2.5-Pro (62.1\%), achieve improved but still limited accuracy, with GPT-4.1 reaching 66.1\%. When integrated with the proposed measure-wise OMR (M-OMR) module, MuseAgent attains 74.1\% accuracy, outperforming both generalist models and music-specialized baselines such as NotaGPT (68.1\%).

\begin{figure*}[!htbp]
\centering

\begin{minipage}[t]{0.4\textwidth}
    \centering
    \caption{Image-to-ABC Conversion Comparison. Results are evaluated on the standardized benchmark introduced in NotaGPT~\cite{tang2025notamultimodalmusicnotation}.}
    \label{tab:convert-abc}
    \resizebox{\textwidth}{!}{
    \begin{tabular}{l|c}
    \toprule
    \textbf{Model} & \textbf{Levenshtein Distance} \\
    \midrule
    VisualGLM-6B & 643.72 \\
    DeepSeek-VL-7B-Chat & 308.27 \\
    LLaVA-v1.5-13B & 147.47 \\
    LLaVA-v1.6-Vicuna-13B & 918.94 \\
    Qwen-VL & 439.82 \\
    NotaGPT-7B & 59.47 \\
    Gemini-pro-vision & 354.30 \\
    GPT-4V & 655.45 \\
    \rowcolor{gray!20} M-OMR (ours) & \textbf{18.39} \\
    \bottomrule
    \end{tabular}}
\end{minipage}%
\hfill
\begin{minipage}[t]{0.57\textwidth}
    \centering
    \caption{Comparisons of open-source models and API-based models.}
    \label{table:comparison}
    \resizebox{\textwidth}{!}{
    \begin{tabular}{lccccc}
    \toprule
    \textbf{Model} & \textbf{LSA} & \textbf{ROUGE-1} & \textbf{ROUGE-L} & \textbf{METEOR} & \textbf{Avg} \\
    \midrule
    InternVL-Chat-v1.5 & 14.96 & 19.71 & 13.32 & 19.68 & 16.92 \\
    VisualGLM-6B & 10.36 & 21.61 & 13.21 & 18.19 & 15.84 \\
    DeepSeek-VL-7B-base & 9.92 & 16.43 & 11.60 & 13.81 & 12.94 \\
    InstructBLIP-Vicuna-7B & 8.28 & 22.23 & 14.93 & 16.74 & 15.55 \\
    InstructBLIP-Vicuna-13B & 8.37 & 20.29 & 14.18 & 14.17 & 14.25 \\
    Qwen-VL & 9.58 & 15.21 & 10.37 & 12.56 & 11.93 \\
    Qwen-VL-Chat & 9.66 & 16.80 & 11.37 & 14.42 & 13.06 \\
    NotaGPT-7B & 12.46 & 22.63 & 15.53 & 18.34 & 17.24 \\
    Gemini-pro-vision & 15.88 & 22.21 & 15.09 & \textbf{20.31} & 18.37 \\
    GPT-4V & 14.03 & 18.49 & 11.36 & 19.94 & 15.96 \\
    GPT-4o & \textbf{15.92} & 18.27 & 11.35 & 20.26 & 16.45 \\
    \rowcolor{gray!20} MuseAgent (w/ M-OMR) & 15.75 & \textbf{24.92} & \textbf{15.76} & 20.17 & \textbf{19.15} \\
    \bottomrule
    \end{tabular}}
\end{minipage}
\vspace{-10px}
\end{figure*}

\subsection{Performance Evaluation of M-OMR Against VLMs}

To isolate the contribution of structured visual perception within MuseAgent, we further evaluate the proposed M-OMR module against state-of-the-art visual language models (VLMs) on music score understanding tasks. Following the evaluation protocol of NotaGPT~\cite{tang2025notamultimodalmusicnotation}, we consider two settings: 
(i) \textbf{closed-set} conversion of sheet music into ABC notation, evaluated using Levenshtein Distance, and 
(ii) \textbf{open-set} visual music analysis, assessed with semantic metrics including LSA, ROUGE, and METEOR. 
This controlled comparison allows us to assess whether measure-wise symbolic grounding provides advantages over monolithic vision–language modeling for structured music understanding.

\subsubsection{Closed-set Image-to-ABC Notation}
We evaluate eight representative MLLMs, including API-based models (e.g., GPT-4V, Gemini Pro) and open-source models (e.g., LLaVA, VisualGLM, Qwen-VL-32B, NotaGPT-7B). As shown in Table~\ref{tab:convert-abc}, M-OMR achieves the lowest Levenshtein Distance (18.39), far surpassing all baselines such as NotaGPT-7B (59.47) and LLaVA-13B (147.47). These results demonstrate M-OMR’s superior structural accuracy in symbolic notation conversion, highlighting its robustness in closed-set tasks.

\subsubsection{Open-set Score Understanding}
For open-set tasks, we compare models on semantic similarity and content relevance (Table~\ref{table:comparison}). MuseAgent with M-OMR achieves the best average score (19.15), outperforming both strong API baselines such as GPT-4o (16.45) and Gemini Pro (18.37), as well as open-source vision–language models. Gains are consistent across metrics: higher ROUGE-1 and METEOR reflect better content coverage and fluency, while improved LSA highlights M-OMR’s ability to capture nuanced musical semantics. Together, these results establish M-OMR as a robust and reliable solution for score interpretation within MuseAgent.

\section{Conclusion}
We introduced \textbf{MuseBench}, a comprehensive benchmark for multimodal music understanding, and \textbf{MuseAgent}, a modular agent that integrates symbolic score parsing and performance audio transcription. MuseBench spans 28 tasks across theory, score, and performance dimensions, offering a rigorous testbed for evaluating the reasoning capabilities of MLLMs. Experiments show that while general-purpose LLMs perform strongly on text-based tasks, they struggle with fine-grained score and audio understanding. By incorporating modality-specific modules such as M-OMR and AMT, MuseAgent achieves substantial gains in both image and audio modalities, demonstrating the necessity of domain-aware perceptual front-ends. These findings highlight the limits of pure scaling in generalist models and confirm the effectiveness of modular integration for complex music reasoning. We hope this work establishes a foundation for future research in AI-assisted music analysis, composition, and education, and for extending multimodal benchmarks beyond text, vision, and speech into the rich domain of music.

\section*{Limitations}
Our benchmark and model currently focus on piano-related tasks, a deliberate design choice motivated by the availability of large-scale data and the standardized nature of piano notation. While this enables controlled evaluation, it limits direct coverage of other instruments and notational systems. Nevertheless, both MuseBench and MuseAgent are designed to be extensible to other domains given appropriate task-specific data.

\bibliography{latex/acl_latex}

\appendix

\section{Appendix}
\label{sec:appendix}

\section{Dataset Details}
\label{app:data_details}

\subsection{Data Sources and Selection Criteria}
\label{app:data_details}

To ensure both diversity and legal compliance, we selected music scores that are either (i) public domain works (composers deceased before 1954), or (ii) explicitly released under open licenses such as Creative Commons. No copyrighted material from the last 75 years was included. All recordings either originated from public-domain datasets or were contributed by professional musicians with signed consent agreements.

The dataset includes:
\begin{itemize}
    \item \textbf{Sheet Music Images:} 
    \begin{itemize}
        \item \textbf{MuseScore}~\cite{musescore_url}: $\sim$300 community-created scores (released under CC licenses), including modern and popular music.
        \item \textbf{IMSLP}~\cite{imslp_url}: $\sim$200 high-resolution classical scores spanning 1600–1920, guaranteed public domain.
        \item \textbf{Mutopia Project \& Project Gutenberg}~\cite{mutopia_url}: $\sim$100 scores with public licenses, covering canonical works by Bach, Mozart, Beethoven, Chopin.
    \end{itemize}
    
    \item \textbf{Textual Descriptions:} Metadata for each score includes title, composer, key, meter, and rhythmic structure, automatically extracted from symbolic files and verified by expert annotators.

    \item \textbf{Audio Files:} 513 piano performance recordings across classical, modern, and popular genres. All recordings are either public-domain (older archive sources) or provided directly by performers under Creative Commons licenses.
\end{itemize}

\subsection{Benchmark License and Usage}

All components of \textbf{MuseBench} are released under strict legal and ethical compliance:

\begin{itemize}
    \item \textbf{Sheet music} is drawn exclusively from public-domain repositories (IMSLP, Mutopia, Project Gutenberg) or from MuseScore where contributors licensed works under Creative Commons. Only works by composers deceased before 1954 are included.
    \item \textbf{Audio recordings} originate from public-domain archives or were directly contributed by professional musicians under written consent and Creative Commons licenses. No copyrighted recordings from the last 75 years are included.
    \item \textbf{Annotations} (ABC, MusicXML, task prompts) were prepared by trained musicians. Annotators provided informed consent, and inter-annotator agreement reached $\kappa=0.87$.
\end{itemize}

\noindent
\textbf{License:} MuseBench is released for non-commercial research and educational purposes under the \textbf{CC BY-NC 4.0} license. Redistribution or reuse of individual scores or recordings must comply with the original source licenses. Upon acceptance, we will publicly release all data, annotations, and evaluation scripts.

\section{Detailed Benchmark Creation}
\label{appendix:task-details}

\subsection{Detailed Task Definition}
\paragraph{Music Theory Understanding.}  
This dimension focuses on textual comprehension of symbolic and conceptual music knowledge. It includes two sub-tasks:
\begin{itemize}
    \item \textbf{Music Theory Recognition:} Evaluates understanding of basic music theory concepts, including key signatures, time signatures, note durations, and rhythmic structures.
    \item \textbf{Music Theory Reasoning:} Involves inferential questions that require deeper reasoning over symbolic descriptions of music, such as determining harmonic progression or identifying musical forms.
\end{itemize}

\paragraph{Sheet Music Understanding.}  
This dimension assesses the model’s ability to interpret notated music from sheet images, and includes:
\begin{itemize}
    \item \textbf{Information Extraction:} Transcription of basic musical metadata such as clefs, key signatures, and tempo markings from visual inputs.
    \item \textbf{Symbolic Analysis:} Understanding note symbols, their spatial and rhythmic relationships, and staff-based structural elements.
    \item \textbf{High-Level Interpretation:} Analyzing expressive or stylistic cues, such as articulation, phrasing, and functional roles in the musical context.
\end{itemize}

\paragraph{Performance Audio Analysis.}  
This dimension assesses the model’s ability to analyze expressive and structural characteristics in real performance recordings. It includes:
\begin{itemize}
    \item \textbf{Performance Evaluation:} Judging the accuracy and completeness of a musical performance, including rhythmic precision, dynamic variation, and articulation clarity.
    \item \textbf{Consistency Evaluation:} Analyzing temporal stability, pitch consistency, and smoothness in expressive transitions across the performance.
\end{itemize}

\subsection{Generation}

Based on the annotated metadata, we constructed multimodal question–answer pairs. Candidate questions were first generated using a combination of rule-based templates and GPT-4o~\cite{openai2024gpt4omini} prompting to increase linguistic variety and naturalness. Ground-truth answers were derived deterministically from symbolic metadata through retrieval, calculation, and statistics, ensuring reproducibility. Expert musicians then reviewed and refined all question–answer pairs to guarantee correctness, clarity, and balanced difficulty. 

The resulting dataset covers text, image, and audio modalities, offering comprehensive multimodal evaluation. Unlike prior datasets such as MusicTheoryBench~\cite{yuan2024chatmusician}, which focus exclusively on text-based symbolic theory questions and include only a few hundred examples, MuseBench introduces multimodal QA tasks that demand joint reasoning over scores, audio, and metadata. This process yields a large-scale benchmark comprising 591 sheet music images, 513 audio recordings, and 2,052 expert-verified question–answer pairs.

\subsection{Dataset Compliance and Licensing}
To ensure ethical and legal compliance, all components of MuseBench are sourced from either public-domain repositories (e.g., IMSLP, Mutopia, Project Gutenberg) or Creative Commons–licensed platforms (e.g., MuseScore), ensuring full legal compliance. Performance recordings are either public-domain or contributed with explicit consent under CC licenses. Further details on data sources, selection criteria, and license terms are provided in Appendix~\ref{app:data_details}.
\begin{table*}[t]
  \caption{Defined tasks in MuseBench. The question format is randomly selected from a format pool for each task. The question types “MCQ,” and “T/F” represent multiple-choice questions and judge true or false.}
  \label{tab:music_taxonomy}
  \centering
  \small
  \resizebox{\textwidth}{!}{%
  \begin{tabular}{@{}lllp{9.5cm}c@{}}
    \toprule
    \textbf{Modality} & \textbf{Ability Dimension} & \textbf{Sub-task} & \textbf{Example Question} & \textbf{Type} \\
    \midrule
    \multirow{8}{*}{\textbf{Text}}
      & \multirow{8}{*}{Music Theory Understanding \& Reasoning}
      & Pitch and Intervals            & How many half steps are present in an augmented sixth interval? & MCQ \\
      & & Rhythm and Meter             & Which time signature represents a compound quadruple meter? & MCQ \\
      & & Harmony and Tonality         & Which pivot chord enables smooth modulation from C major to G major? & MCQ \\
      & & Melody Structure             & In a period structure, what describes the second phrase that resolves the first? & MCQ \\
      & & Scales and Modes             & The Dorian mode starting on D is derived from which major scale? & MCQ \\
      & & Clefs and Notation           & In alto clef, which pitch class is on the fourth space? & MCQ \\
      & & Dynamics and Articulations   & How should a musician perform staccato notes marked with a crescendo? & MCQ \\
      & & Form and Structure           & Which structural pattern best describes sonata-rondo form? & MCQ \\
    \midrule
    \multirow{16}{*}{\textbf{Image}\vspace{-150pt}} 
      & \multirow{6}{*}{Information Extraction}
      & Title Recognition             & What is the title of the piece? & MCQ \\
      & & Composer Recognition          & Who is credited as the composer of the piece titled ``Classical Rag''? & MCQ \\
      & & Meter Recognition             & What is the meter signature of the piece titled ``The Waltz on my bum''? & MCQ \\
      & & Key Signature Recognition     & What is the key signature of the piece at the beginning of the score? & MCQ \\
      & & Voice/Instrument Identification & Which voices are assigned to the bass clef? & MCQ \\
    \cmidrule{2-5}
      & \multirow{5}{*}{Symbolic Analysis\vspace{-38pt}}
      & Note Localization             & In which measure does voice V:1 first play a chord containing the note E natural above middle C? & MCQ \\
      & & Note Frequency Statistics     & In voice V:3, which pitch class appears most frequently as a sounding note (excluding rests and grace notes)? & MCQ \\
      & & Rhythmic Pattern Analysis     & In the first four measures of the melody line (V:1), which rhythmic pattern is predominantly used? & MCQ \\
      & & Harmony Identification        & In measure 18, which chord is formed by the soprano (V:1) and bass (V:3)? & MCQ \\
      & & Instrumentation               & Which staves in this score are written in the bass clef? & MCQ \\
    \cmidrule{2-5}
      & \multirow{6}{*}{High-Level Musical Understanding\vspace{-50pt}}
      & Functional Harmony Analysis   & What is the harmonic function of the raised A note (\^a) in E-flat major? & MCQ \\
      & & Mode Identification           & Considering key signature and accidentals, which mode is implied? & MCQ \\
      & & Melodic Range                 & What is the melodic range of voice V:1? & MCQ \\
      & & Ornamentation Recognition     & Which ornamentation is most consistently applied? & MCQ \\
      & & Non-Chord Tone Identification & In measure 2 of Voice 1, which non-chord tone acts as a passing tone? & MCQ \\
      & & Key Modulation Recognition    & At which measure does modulation from E-flat major occur? & MCQ \\
    \midrule
    \multirow{4}{*}{\textbf{Audio}}
      & \multirow{2}{*}{Performance Evaluation}
      & Key Accuracy     & Measure~11 confirms full key accuracy. & T/F \\
      & & Completeness    & Measure~8 has no missing notes. & T/F \\
    \cmidrule{2-5}
      & \multirow{2}{*}{Consistency Evaluation}
      & Tempo Stability  & Measure~11’s tempo matches reference stability. & T/F \\
      & & Speed           & Measure~8's speed is significantly slower than required. & T/F \\
    \bottomrule
  \end{tabular}}
\end{table*}

By integrating tasks across text, image, and audio modalities, the \textbf{MuseBench} dataset offers a comprehensive evaluation platform for multimodal large language models, spanning every facet of music understanding—from music theory comprehension to the assessment of real performance characteristics.

\section{Implementation Details of the M-OMR}
\label{appendix:omr-examples}

The M-OMR module bases on a “divide-and-combine” strategy that serves as a visual encoder specialized for music score.

\paragraph{Divide.}
The input score image is initially segmented into individual measures through a combination of staff line detection and barline localization. Each segmented measure is then treated as an independent visual unit for localized recognition. Specifically, a YOLOv8-based detector \cite{varghese2024yolov8} is employed to identify and localize each measure, 

\paragraph{Process.}
Each measured image is encoded into a high-dimensional embedding using a ResNet-50 backbone ~\cite{he2016deep}, capturing fine-grained visual features of musical symbols, including clefs, staves, barlines, key signatures, and time signatures. These embeddings are then sequentially decoded into ABC-format symbolic sequences using an LSTM-based decoder \cite{yu2019review} trained for note-level transcription. 

\paragraph{Combine.}
The measure-level symbolic sequences are aggregated to reconstruct the full musical piece. During this step, time signatures extracted during pre-processing are aligned with each measure to ensure consistent rhythmic context. The final output is a well-formed ABC representation that preserves both temporal structure and notational correctness.

Recent studies have demonstrated the effectiveness of YOLO-based models in structured document analysis tasks \cite{zhao2024doclayout}. 

Then, each segmented measure image is passed through a ResNet-50 ~\cite{he2016deep} encoder to obtain a latent visual embedding $\mathbf{x}_t$. The decoder is implemented as a unidirectional LSTM, which autoregressively generates the corresponding ABC sequence token-by-token.

\begin{equation}
\mathbf{h}_t = \text{LSTM}(\mathbf{h}_{t-1}, \mathbf{x}_t; \theta),
\end{equation}

After decoding all measures, the symbolic output is reconstructed via:

\begin{equation}
\text{ABC}_{\text{full}} = \text{ConcatMeasures}\left\{ (\text{ABC}_i, \text{TimeSig}_i) \right\}_{i=1}^{n},
\end{equation}

where $\text{TimeSig}_i$ is the pre-detected time signature of measure $i$, and $n$ is the total number of measures.

\paragraph{Datasets.} To construct a robust optical music recognition (OMR) module, we curated a large-scale dataset derived from the MuseScore platform, comprising over 80,000 music scores. Each score was first converted from MusicXML format to ABC notation~\cite{walshaw2021abc}, and subsequently rendered into SVG images. To further increase data diversity and model robustness, we performed structured data augmentation by randomly shuffling and replacing ABC bars, resulting in a synthetic corpus of 2.3 million ABC samples. Following image generation, we employed YOLO-based~\cite{varghese2024yolov8} segmentation to automatically detect and extract individual bars from the SVGs, ultimately yielding over \textbf{10 million} of image-bar pairs.

\paragraph{Training Configurations.}  The training was conducted over 100 epochs using a batch size of 12 and a learining rate of 1e-4. Our model achieved near accuracy (approximately 98\%) on our held-out validation set, demonstrating both the scale and effectiveness of our training pipeline.

\begin{figure*}[htbp]
  \centering
  \includegraphics[width=\linewidth]{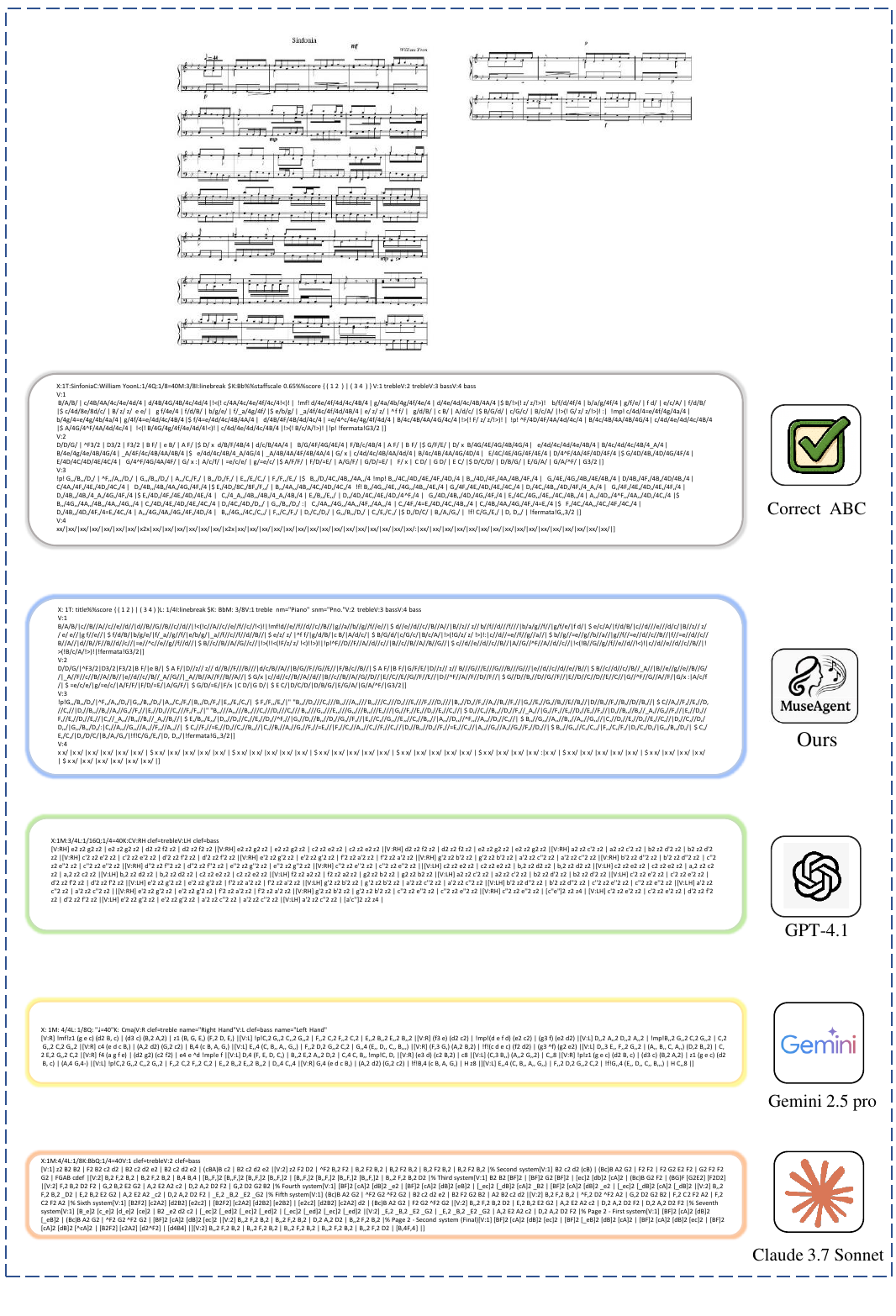}  
  \caption{Performance of different LLMs on converting sheet music images to ABC notation}
  \label{fig:appendix-abc}
\end{figure*}

The visualization samples of ABC notation can be found in Figure~\ref{fig:appendix-abc}. From the figure, we observe that MuseAgent, equipped with the M-OMR module, is able to accurately transcribe the entire sheet music into ABC notation. In contrast, other large language models struggle to extract complete and precise ABC representations, often missing structural or symbolic details.

\section{Implementation Details of AMT and Audio-to-Score Alignment}
\label{appendix:amt}

This appendix provides the detailed implementation of the Automatic Music Transcription (AMT) and Audio-to-Score Alignment modules used in MuseAgent. While we adopt methods inspired by prior work~\cite{hawthorne2017onsets,nakamura2015real}, we report all relevant architecture and configuration details to facilitate reproducibility and downstream integration.

\subsection{Automatic Music Transcription (AMT)}

\paragraph{Input Representation.}
We use the constant-Q transform (CQT) to extract time-frequency features from raw audio. A CNN–BiLSTM \cite{siami2019performance} architecture predicts both onset and frame-level note activations, following the structure of Onsets-and-Frames~\cite{hawthorne2017onsets}. 

Formally, given input features $\mathbf{X}_t$, the onset probability is predicted as:
\begin{equation}
\mathbf{O}_t = \sigma(\text{BiLSTM}(\text{CNN}(\mathbf{X}_t))),
\end{equation}
and the framewise activation is computed as:
\begin{equation}
\mathbf{F}_t = \sigma(\text{BiLSTM}([\text{CNN}(\mathbf{X}_t), \mathbf{O}_t])).
\end{equation}

The parameters are listed in Table~\ref{tab:cqt}.

\begin{table}[h]
\centering
\caption{CQT Configuration for AMT}
\label{tab:cqt}
\begin{tabular}{lc}
\toprule
Parameter & Value \\
\midrule
Sample Rate & 16 kHz \\
Hop Length & 512 samples \\
Frequency Bins & 88 (covering A0–C8) \\
Bins per Octave & 12 \\
Window Function & Hann \\
Normalization & Log-magnitude \\
\bottomrule
\end{tabular}
\end{table}

\paragraph{Network Architecture.}
The AMT model processes the CQT input through:
\begin{itemize}
    \item \textbf{CNN Frontend:} 3 convolutional layers (kernel size: $3\times3$, stride: 1, padding: 1), each followed by ReLU and batch normalization.
    \item \textbf{BiLSTM Layer:} One bidirectional LSTM with 128 hidden units per direction.
    \item \textbf{Onset Head:} Fully connected layer with sigmoid activation to predict per-frame note onsets.
    \item \textbf{Frame Head:} Similar layer conditioned on onset features, used to predict framewise note activations.
\end{itemize}

\paragraph{Training Details.}
\begin{itemize}
    \item \textbf{Loss:} Binary cross-entropy loss applied independently to onset and frame predictions.
    \item \textbf{Optimizer:} Adam with learning rate $1\times10^{-4}$.
    \item \textbf{Training Epochs:} 50 on MAESTRO-V3~\cite{hawthorne2018enabling}.
    \item \textbf{Batch Size:} 8.
\end{itemize}

\paragraph{Post-Processing.}
Binary predictions are thresholded at 0.5. Onset and frame activations are merged into MIDI note events using the following heuristic:
- A note onset is declared if the onset activation exceeds the threshold.
- Note duration is extended over consecutive frames with active predictions.

The final output is exported in either MIDI or MusicXML format. We further convert to ABC notation when needed for symbolic alignment.

\subsection{Audio-to-Score Alignment}

\paragraph{Model Overview.}
We adopt the dual-layer HMM approach from~\cite{nakamura2015real}, which allows robust alignment between symbolic scores and AMT-derived audio events. The alignment process maximizes the posterior probability of the score position $p_t$ given observed acoustic features $x_t$:
\begin{equation}
p(p_t \mid x_t) = \frac{p(x_t \mid p_t) \cdot p(p_t)}{p(x_t)}.
\end{equation}

\paragraph{Structure.}
\begin{itemize}
    \item \textbf{Top-layer HMM:} Models transitions between score positions (e.g., measures or note groups).
    \item \textbf{Bottom-layer HMM:} Captures fine-grained temporal dynamics within a note (onset, sustain, silence).
\end{itemize}

\paragraph{Observation Model.}
The likelihood $p(x_t \mid p_t)$ of observing acoustic feature $x_t$ given score position $p_t$ is modeled by a Gaussian Mixture Model (GMM):
\begin{itemize}
    \item Number of components: 8
    \item Covariance: Diagonal
    \item Input: PCA-reduced CQT (dimension = 30)
    \item Training: Expectation-Maximization on aligned score–audio pairs
\end{itemize}

\paragraph{Transition Model.}
We define a transition matrix $A$ that supports:
\begin{itemize}
    \item \textbf{Self-loop:} Sustains the current note position.
    \item \textbf{Forward transition:} Normal sequential progression.
    \item \textbf{Backward jump:} Repeat sections or corrections.
    \item \textbf{Forward skip:} Skipping sections.
\end{itemize}

These transitions are encoded as probabilities:
\[
A_{ij} = p(p_t = j \mid p_{t-1} = i), \quad \text{with non-zero mass for } |i - j| > 1.
\]

\paragraph{Inference.}
We use Viterbi decoding to compute the most probable alignment path:
\[
p_{1:T}^* = \arg\max_{p_{1:T}} \prod_{t=1}^{T} p(x_t \mid p_t) \cdot A_{p_{t-1}, p_t}
\]

This algorithm effectively handles expressive timing, omission, repetition, and incorrect notes, making it robust for alignment in real-world performance scenarios.

\subsection{Evaluation Results of AMT and Alignment Algorithms}

This section presents the evaluation results of the Automatic Music Transcription (AMT) and alignment algorithms, which are recorded in a JSON format. These results provide a detailed assessment of various transcription metrics, such as overall accuracy, note matching, speed, stability, and tempo synchronization.

The evaluation results for measure 37 of a random sample are summarized in the table below:

\begin{table}[htbp]
\centering
\tiny
\begin{tabular}{l|l}
\hline
\textbf{Metric}           & \textbf{Value}                  \\ \hline
Overall Evaluation (eva\_all)        & 0.9252619743347168             \\ \hline
Note Evaluation (eva\_note)          & 1.0                             \\ \hline
Speed Evaluation (eva\_speed)        & 1.0                             \\ \hline
Stability Evaluation (eva\_stability)  & 0.7282252907752991          \\ \hline
Tempo Synchronization (eva\_tempo\_sync) & 1.0                            \\ \hline
Extra Notes Count        & 0                               \\ \hline
Matched Notes Count      & 2                               \\ \hline
Missing Notes Count      & 0                               \\ \hline
\end{tabular}
\caption{Evaluation Results for Measure 37 of a Random Sample}
\end{table}

Additionally, the following figure~\ref{fig:json} provides a visual representation of the performance comparison across the various transcription metrics.

\begin{figure*}[htbp]
\centering
\includegraphics[width=0.95\textwidth]{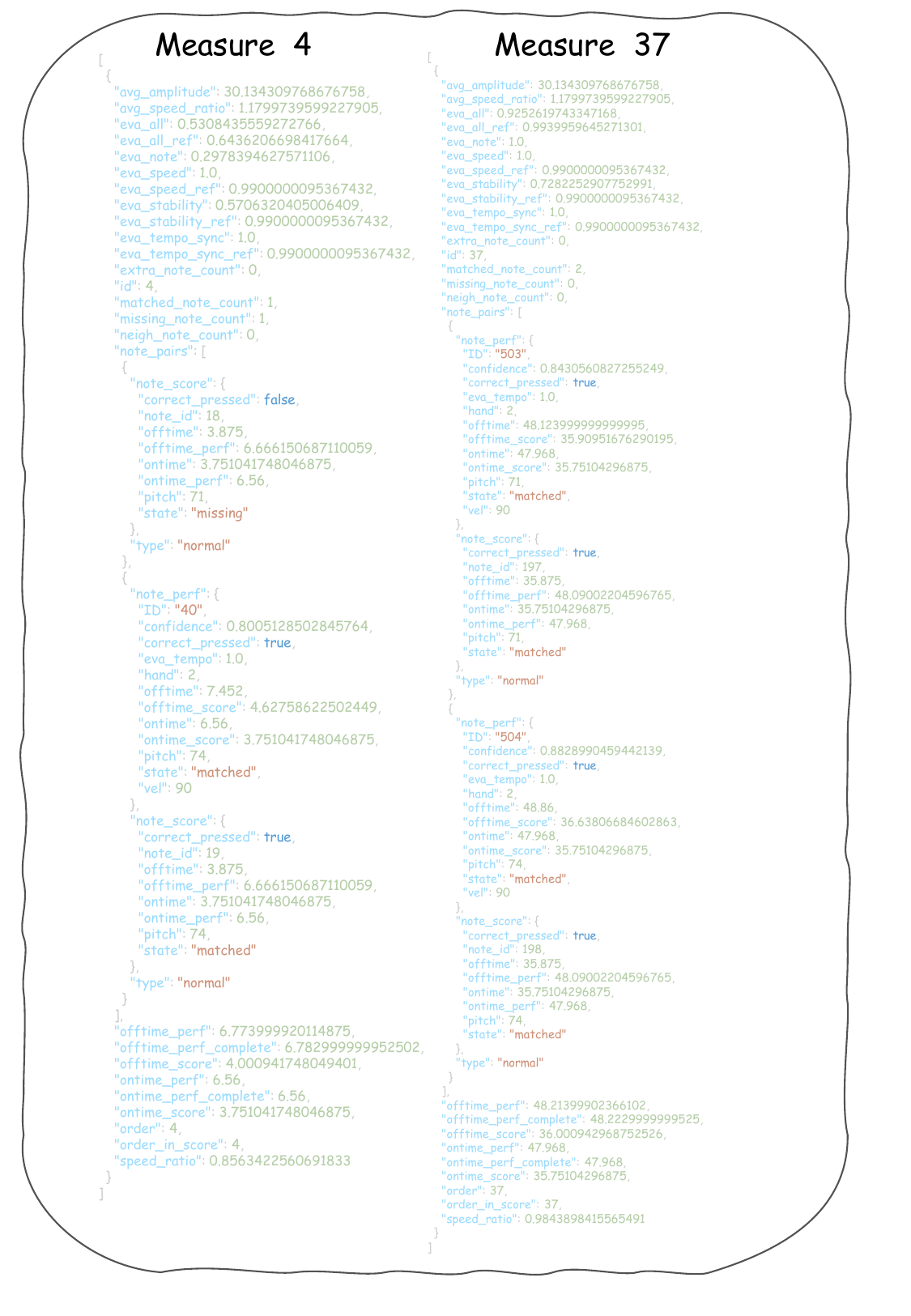}
\caption{Results of Performance Comparison Across Transcription Metrics}
\label{fig:json}
\end{figure*}

\section{Evaluation Details}
\label{appendix:evaluation}
\subsection{Evaluation Details in our Benchmark}

We present the different prompts used for three modalities: text, image, and audio. The following table summarizes the specific prompts for each modality.

\begin{table*}[H]
    \centering
    \caption{Prompts used for evaluation in our benchmark. The <\texttt{measure\_id}> represents the unique identifier for each musical measure or section in question.}
    \resizebox{\textwidth}{!}{%
    \begin{tabular}{c|c}
    \hline
    \textbf{Modality} & \textbf{System Prompt} \\ \hline
    Text & You are a music expert. Please read the following question carefully and provide the correct answer based on your knowledge of music theory and practice. \\ \hline
    Image & You are a music expert. Please analyze the given sheet music image and select the correct answer to the question based on its notated content. \\ \hline
    Audio & You are a music expert. Please carefully listen to the <\texttt{measure\_id}> section of the provided audio excerpt and answer the question based on your auditory analysis. \\ \hline
    \end{tabular}%
    }
\end{table*}

\subsection{Evaluation Metrics Used in Contrast Experiment}

In this appendix, we present the evaluation metrics used in our M-OMR, comparing it with different models for converting images to ABC notation text, utilizing levenshtein distance. Additionally, we analyze music content using semantic similarity and word Matching metrics.

\paragraph{Levenshtein Distance.}The Levenshtein Distance~\cite{yujian2007normalized} is used as the evaluation metric for converting images to ABC notation text. It refers to the minimum number of single-character operations required to transform model responses into the correct answer sequence. 

Let \( D \) be a matrix of size \( (|R| + 1) \times (|A| + 1) \), where \( |R| \) and \( |A| \) represent the lengths of the response and answer sequences, respectively. \( D[i][j] \) denotes the minimum edit distance between the first \( i \) characters of \( R \) and the first \( j \) characters of \( A \).

The subsequent values of \( D \) are computed using the following recurrence relation:

\[
D[i][j] = \min \left\{
\begin{array}{ll}
D[i-1][j] + 1 & \text{(delete)} \\
D[i][j-1] + 1 & \text{(insert)} \\
D[i-1][j-1] + \text{cost} & \text{(substitute)}
\end{array}
\right.
\]

where the cost is 0 if \( R[i-1] = A[j-1] \), otherwise it is 1.

\paragraph{Semantic Similarity and Word Matching Metrics.}Our Experiment also uses two categories of metrics: semantic similarity and word matching, for analyzing music content.

For semantic similarity, we use Latent Semantic Analysis (LSA), which measures the semantic similarity of text by computing the cosine similarity between vectors. The cosine similarity is given by:



For word matching, we use the following metrics:

\begin{itemize}
    \item \textbf{ROUGE-1}: Calculates the number of unigram matches between the generated and reference text.
    \item \textbf{ROUGE-L}: Measures the longest common subsequence (LCS) match between the generated and reference text.
    \item \textbf{METEOR}: Calculates synonym matches and uses a combination of unigram matches, longest common subsequences, and synonym matches.
\end{itemize}

\end{document}